\documentclass[preprint,12pt]{elsarticle}

\usepackage{amssymb}
\usepackage{amsmath}

\journal{Journal of computation of physics}

\begin{document}

\begin{frontmatter}



\title{General wetting energy boundary condition in a fully explicit non-ideal fluids solver }


\author[a]{Chunheng Zhao}

\author[a]{Alexandre Limare}
\author[a,b]{Stephane Zaleski}

\address[a]{Sorbonne Universit\'{e} and CNRS, Institut Jean Le Rond d'Alembert UMR 7190, F-75005 Paris, France}
\address[b]{Institut Universitaire de France, Paris, France}

\begin{abstract}

We present an explicit finite difference method to simulate the non-ideal multi-phase fluid flow. The local density and the momentum transport are modeled by the Navier-Stokes (N-S) equations and the pressure is computed by the Van der Waals equation of the state (EOS). The static droplet and the dynamics of liquid-vapor separation simulations are performed as validations of this numerical scheme. In particular, to maintain the thermodynamic consistency, we propose a general wetting energy boundary condition at the contact line between fluids and the solid boundary. We conduct a series of comparisons between the current boundary condition and the constant contact angle boundary condition as well as the stress-balanced boundary condition. This boundary condition alleviates the instability induced by the constant contact angle boundary condition at $\theta \approx0 $ and $\theta \approx \pi$. Using this boundary condition, the equilibrium contact angle is correctly recovered and the contact line dynamics are consistent with the simulation by applying a stress-balanced boundary condition. Nevertheless, unlike the stress-balanced boundary condition for which we need to further introduce the interface thickness parameter, the current boundary condition implicitly incorporates the interface thickness information into the wetting energy.
\end{abstract}

\begin{highlights}
\item Energy consistent boundary condition for single species multi-phase Van der Waals' model
\item Explicit finite difference method with adaptive mesh refinement
\end{highlights}


\end{frontmatter}


\section{Introduction}
Fluids spreading on solids are practical multi-phase systems in the real world~\cite{de1985wetting}. Industrial applications of solid wetting research can be found in 3D printing \cite{shahrubudin2019overview}, nucleate boiling~\cite{pioro2004nucleate,jones2009influence}, and surface material construction~\cite{koch2008diversity,bhushan2007wetting}. Numerical simulations on the wetting problem are made extremely difficult or impossible by the existence of a very wide range from the macroscopic scales to the nanometric scales~\cite{cox1986dynamics,koroteev2014direct,lacis2020steady,lacis2022nanoscale}. When it comes to models such as level-set and volume of fluid (VOF) that treat the interface between two fluids as a sharp interface, the no-slip boundary condition contradicts the actual behavior observed in droplet spreading~\cite{lacis2022nanoscale}. To address the limitation of those methods on the moving contact line, researchers have implemented explicit Navier-slip or implicit numerical slip boundary conditions~\cite{navier1823memoire,spelt2005level,afkhami2009mesh}.  Nevertheless, the boundary conditions associated with the sharp interface method will introduce nonphysical dynamics and prove ineffective in handling small or large contact angles. Hence, it is worth considering the diffuse interface method, a thermodynamically consistent mathematical model for multi-phase systems, to effectively simulate the dynamics of contact lines~\cite{seppecher1996moving,jacqmin2000contact,yue2010sharp,lee2010lattice,zhu2019thermodynamically,jamet2001second}. The diffuse interface method introduces energy dissipation, enabling the modeling of droplet spreading even with the no-slip or small slip length boundary condition~\cite{jacqmin2000contact,yue2010sharp,yue2020thermodynamically}. In the vicinity of a diffused contact line, the bulk free energy and the surface energy determine the contact line profile as well as the fluid flow~\cite{seppecher1996moving,seppecher1996moving,cahn1958free,allen1979microscopic,safari2014consistent}. Moreover, the boundary condition within the diffuse interface method can be described as the wetting energy to ensure the thermodynamic consistency~\cite{jacqmin1996energy,jones2009influence,jacqmin2000contact,yue2010sharp,lee2010lattice,zhao2023engulfment}. By employing the diffuse interface method, it becomes possible to accurately simulate contact angles, regardless of whether they are small or large in magnitude.

A well-known classical diffuse interface method is derived from the Van der Waals (VDW) equation of state (EOS) for a single species, $\left(p+a\rho^2\right)(1/\rho-b)=RT$, with classical notations, where $a$ and $b$ are modification parameters of the molecular interaction and the molecular volume respectively~\cite{van1979thermodynamic,margenau1939van,dzyaloshinskii1961general}. Under the pressure and energy-driven mechanism, the VDW method is able to separate the single species into two phases, one with higher density, and the other one with lower density. Compared to the Cahn-Hilliard (C-H) method, the VDW has some different characteristics to be noted. First, the VDW method describes the single species phase change where the interface is indicated by the local density $\rho$, while the C-H method describes the physical situation of a binary system of two essentially immiscible species and the interface profile is normally steeper than the VDW method. In addition, in the VDW model, the bulk energy density is computed by the entropy and molecular interaction which can be represented by the equation $\rho f_0=- \rho RT\log{\left(1/\rho-b\right)}-a\rho^2$. In contrast, the bulk free energy density adopted in the C-H is a double-well fourth-order polynomial $\rho f_0=\beta (\rho-\rho_l)^2(\rho-\rho_g)^2$, where $\beta$ denotes the constant bulk energy coefficient, and $\rho_l$, $\rho_g$ are saturated liquid and gas densities. One of the benefits of using the C-H type energy form is it allows us to accurately represent the flat interface profile at equilibrium using a hyperbolic tangent function. Additionally, the C-H energy form enables us to explicitly determine the interface thickness and the surface tension~\cite{chen2023diffuse}. However, the inclusion of a fourth-order partial differential equation greatly amplifies the intricacy of the problem, thereby intensifying the difficulty of numerically simulating the C-H equation. Conversely, the VDW method offers a viable diffuse interface approach that is not only comparatively efficient but also valid.

Over the past few decades, extensive research has been conducted to numerically investigate the diffuse interface model of single species multi-phase systems~\cite{nadiga1995investigations,lee2006eliminating,liu2009wall,huang2022surface}, and various boundary condition methods have been employed in the context of the diffuse interface model~\cite{jacqmin2000contact,qian2006variational,yue2010sharp,lee2008wall}. The stress-balanced boundary condition, as proposed in \cite{qian2006variational}, takes into account a smooth variation of surface tension at the diffused interface along the solid boundary. Moreover, from a thermodynamic perspective, the energy-consistent boundary condition can be applied in the diffuse interface method~\cite{jacqmin1996energy, yue2004diffuse,briant2004lattice,briant2004lattice1,liu2009wall}. It establishes a connection between the bulk free energy and the wetting energy at the boundary, ensuring a uniform interface thickness as the system reaches thermodynamic equilibrium. The above-mentioned boundary conditions are based on the C-H type bulk free energy formulation. In this case, 
the wetting energy and the surface tension can be evaluated without the difficulty to compute the integral operation. However, as for the VDW energy form, the value of the interface thickness and the surface tension is not explicit. In order to obtain the surface tension, we need to further numerically compute the integral along the surface, which makes it challenging to apply the mentioned boundary conditions. In recent years, constant contact angle boundary condition \cite{ding2007wetting,chen2023diffuse}, and chemical potential based boundary condition have been employed for the VDW single species model \cite{wen2017chemical,huang2023equation}. As we shall see the constant contact angle boundary condition induces an instability at equilibrium contact angles $\theta_{eq}\approx0$ or $\theta_{eq}\approx\pi$. In addition, the boundary condition used in \cite{huang2022surface} is applied to the pseudopotential LBM method. In this approach, the exact determination of the contact angle requires several free parameters, which adds complexity when utilizing other simulation methods.

In this study, as the energy-consistent boundary condition used in C-H model, we provide a general energy-consistent boundary condition for the VDW single species multi-phase model \cite{jacqmin1996energy,yue2004diffuse,liu2009wall,yue2020thermodynamically}. The boundary condition ensures energy consistency and allows for a uniform interface profile as the equilibrium contact angle is approached. To solve the Navier-Stokes equations, which incorporate a Korteweg stress form to model the surface effect, we employ a fully explicit finite-difference method \cite{nadiga1995investigations,maccormack1982numerical,zhao2023interaction}. This finite difference scheme enables easy implementation of adaptive mesh refinement technology which further enhances the computational efficiency of our approach. We perform a comparison of various boundary condition methods and present the wetting energy for different equilibrium contact angles and interface thickness parameters. Furthermore, we validate our numerical scheme by showcasing two benchmark problems: a single static droplet and the dynamics of liquid-vapor separation. The energy evolution for Laplace numbers $La=[10,1000]$ is shown for a single static droplet and the average domain length evolution of the dynamics of liquid-vapor separation is provided. 
\section{Methodology}
\label{sec:methodology}

In this section, we provide an introduction to the mathematical model utilized in this work. We begin by presenting the governing equations and the thermodynamic energy of the system. With a focus on the energy aspect, we derive the wetting energy and proceed to compare different boundary condition methods based on their profiles while simulating a simple one-dimensional (1-D) equilibrium surface.
\subsection{Governing Equations}
The governing equations employed in our study consist of the compressible Navier-Stokes equations, incorporating the Korteweg stress surface tension force, along with the equation of state (EOS)~\cite{van1979thermodynamic}. Those formulations can be expressed as:
\begin{equation}\label{ns_c}
    \frac{\partial\rho}{\partial t}+\nabla\cdot(\rho\mathbf{u})=0,
\end{equation}
\begin{equation}\label{ns_m}
    \frac{\partial\rho\mathbf{u}}{\partial t}+\nabla\cdot(\rho\mathbf{u}\otimes\mathbf{u})=
    \nabla\cdot
    \left(\sigma_{v}+\sigma_{s}-p\mathbf{I}
    \right),
\end{equation}

\begin{equation}\label{eos}
    p=\rho RT\left(\frac{1}{1-b\rho}-\frac{a\rho}{RT}\right).
\end{equation}
Eqs.~(\ref{ns_c}) and (\ref{ns_m}) is the continuity equation and momentum equations. Here, the operator $\otimes$ represents the tensor product operation. Eq.~(\ref{eos}) stands for the VDW's EOS from which we can obtain the pressure and close this non-ideal gas system. In Eq.~(\ref{ns_c}), $\rho$ denotes the local density of the liquid or gas phase, and $\mathbf{u}$ is the velocity vector. In Eq.~(\ref{ns_m}), $p\mathbf{I}$ is the pressure tensor, where $\mathbf{I}$ is the identity matrix,
\begin{equation}
    \sigma_{v}=     \eta\left[\left(\nabla \mathbf{u}+\nabla^T\mathbf{u}\right)-\frac{2}{3}(\nabla\cdot \mathbf{u})\mathbf{I}\right]
\end{equation}
represents the viscous stress tensor, and
\begin{equation}
    \sigma_{s}= \lambda\left[\left(\frac{1}{2}|\nabla\rho|^2+\rho\nabla^2\rho\right)\mathbf{I}-\nabla\rho \otimes\nabla\rho\right]
\end{equation}
is the surface stress tensor. Within these equations, $\eta$ is the local viscosity, while $\lambda$ corresponds to the surface energy coefficient. It should be noted that the thermodynamic pressure denoted by $p$  can be determined from Eq.~(\ref{eos}), where $R$ denotes the universal gas constant, $T$ is defined as the temperature, and $a$, $b$ are two gas constants that signify the intermolecular attraction and the volume modification ratio, respectively. Normally, we can rearrange Eq.~(\ref{eos}) in a dimensionless form:
\begin{equation}
    p'=-\frac{8\rho'T'}{3-\rho'}-3\rho'^2,     
\end{equation}
where $p'=p/p_c$, $\rho'=\rho/\rho_c$, and $T'=T/T_c$ are the dimensionless forms of pressure, density, and temperature. $p_c=\frac{3}{8}\rho_c R T_c$, $\rho_c=\frac{1}{3b}$, and $T_c=\frac{8a}{27Rb}$ are critical pressure, density and temperature. For our simulations, we select the values $a=3$ and $b=1/3$, leading to $\rho_c=1$ and $p_c=1$. As the pressure term solely appears in a derivative form, we calculate $\nabla p'$ during the simulation instead.

An expression for the energy associated with the pressure term can be expressed as follows~\cite{nadiga1995investigations}:
\begin{equation}
    p=\rho^2 \frac{\partial f_0}{\partial \rho},
\end{equation}
where the Helmholtz free energy per unit volume is expressed as $\rho f_0$. The dimensionless formula $f_0'=\rho_c f_0/p_c$ is given by~\cite{liu2015numerical,onuki2007dynamic}:
\begin{equation}
    f_0'=-\frac{8}{3} T'\log(\frac{1}{\rho'}-\frac{1}{3})-3\rho'- \mu^*,
\end{equation}
where $\mu^*$ denotes the dimensionless bulk chemical potential~\cite{rowlinson2013molecular}, which is a universal constant value in both the liquid and gas regions. The bulk chemical potential can be determined through the Maxwell construction of the pressure profile or the common tangent construction of the free energy~\cite{clerk1875dynamical}.

\subsection{Wetting energy model}
The energy derivation presented in~\cite{jacqmin1996energy} establishes a connection between the stress form and potential form surface tension force formulations. In addition, when there is a solid boundary in simulation, a wetting energy, and a constraint function were introduced to close the system. 
As outlined in \cite{ren2011derivation,liu2009wall,yue2020thermodynamically,lee2008wall}, we derive the boundary condition for the VDW model from an energy perspective. To incorporate the surface effect, we introduce a mixed energy density formulation, where the surface energy per unit volume is expressed as follows:
\begin{equation}
    e_{s}=\frac{\lambda}{2}|\nabla\rho|^2,
\end{equation}
and the mixed energy per unit volume is:
\begin{equation}
    e_{mix}=\rho f_0+e_{s}.
\end{equation}
In this expression, we also consider the kinetic energy per unit volume $\rho e_{k}=\frac{1}{2}\rho|\mathbf{u}|^2$ and the wetting energy per unit area $e_w$. The total energy of the system can be expressed in integral form as follows:
\begin{equation}\label{energytotal}
    E=\int_{\Omega} \left(e_{mix}+\rho e_{k}\right) \ d\mathbf{x}+\int_{\partial\Omega}e_w \ ds.
\end{equation}
Considering a constant temperature, viscous dissipation is the only dissipation of the energy. The evolution of the total energy $E$ is then
\begin{equation}\label{energyevolution}
    \frac{\partial E}{\partial t}=\int_\Omega \left(\frac{\partial e_{mix}}{\partial t}+\frac{\partial \rho e_k}{\partial t}\right) d\mathbf{x}
    +\int_{\partial \Omega} \frac{\partial e_w}{\partial t }ds=\int_\Omega\mathbf{u}\cdot\nabla\cdot\sigma_{v}\ d\mathbf{x}.
\end{equation}
In this equation, $\Omega$ represents the fluid-dominated region, while $\partial \Omega$ corresponds to the solid boundary. Through variable substitution and integration by parts, Eq.~(\ref{energyevolution}) can be rearranged as follows:
\begin{align*}\label{arrange}
\int_\Omega \left[\left(\frac{\partial \rho f_{0}}{\partial \rho}+\lambda\nabla\rho \cdot \nabla \right)
\frac{\partial \rho}{\partial t}+\frac{\partial \rho e_k}{\partial t}\right] d\mathbf{x}
    +\int_{\partial \Omega} \frac{\partial e_w}{\partial t }ds&=\\ \int_\Omega\left(\mu_{mix}\frac{\partial \rho}{\partial t}+\frac{\partial \rho e_k}{\partial t}\right)\ d\mathbf{x}+\int_{\partial \Omega} \left(\lambda\partial_{\mathbf{n}}\rho+\frac{\partial e_w}{\partial \rho}\right)\frac{\partial \rho}{\partial t} \ ds,
\end{align*}
where $\mu_{mix}=\delta e_{mix}/\delta \rho$ represents the mixed chemical potential, which is obtained by taking the functional derivative of the mixed energy. To ensure non-dissipation at the boundary, we obtain the following expression:
\begin{equation}\label{intboundary}
    \lambda\partial_\mathbf{n}\rho+\frac{\partial e_w}{\partial \rho}=0,
\end{equation}
where $\partial_\mathbf{n}\rho$ denotes the wall direction derivative of the density, $\partial e_w/\partial \rho$ is referred to as the wetting potential. The potential surface force formulation can be derived from the volume integral part. The consistency between the potential form and the stress-form formulations can be demonstrated through the inclusion of an additional stress term:
\begin{equation}\label{balance}
   \nabla\cdot \left(\sigma_{s}-p\mathbf{I}\right)=-\rho\nabla\mu_{mix}+\nabla\cdot\sigma_\rho,
\end{equation}
where the additional stress term $\sigma_\rho$ takes the form:
\begin{equation}\label{add}
    \sigma_\rho=\lambda\left(|\nabla\rho|^2\mathbf{I}-\nabla\rho\otimes\nabla\rho\right).
\end{equation}
By utilizing the potential surface force formulation, the presence of spurious currents can be significantly reduced to a level below the round-off limit~\cite{lee2006eliminating,jamet2002theory}. 

For a 1-D planar simulation with $\sigma_\rho=0$, in the equilibrium state of the system, the mixed chemical potential must satisfy the following condition:
\begin{equation}\label{inte}
  \mu_{mix}=\frac{\partial \rho f_0}{\partial\rho}-\lambda\frac{d^2\rho}{d x^2}=0.
\end{equation}
When we multiply Eq.~(\ref{inte}) by $d\rho/d x$ and integrate it, the following equation can be obtained:
\begin{equation}
    \frac{\lambda}{2}\left(\frac{d\rho}{dx}\right)^2=\int \frac{\partial \rho f_0}{\partial x}\ dx. 
\end{equation}
The first derivative of the density can then be derived as:
\begin{equation}\label{d_rho}
    \left|\frac{d\rho}{dx}\right|=\sqrt{\frac{2\rho f_0}{\lambda}}.
\end{equation}
To extend Eq.(\ref{d_rho}) to multi-dimensional problems, we make the approximation $|\nabla\rho|\approx\sqrt{2\rho f_0/\lambda}$. Considering the constraint given by Eq.(\ref{intboundary}), we can derive an energy-consistent wetting energy per unit area as follows:
\begin{equation}\label{w_en1}
    e_{w1}=\cos{\theta_{eq}}\int_{\rho_{gs}}^{\rho_{ls}}\sqrt{2\lambda \rho f_0} \ d\rho +C.
\end{equation}
Here, $C$ represents a constant parameter. However, in the simulation, this constant does not affect the evolution of the contact line, so we can set $C=0$. In this case, the wetting potential $e_{w1}$ is characterized by two saturation densities, and these values align with the equilibrium density in the liquid and gas phases~\cite{liu2009wall}. Therefore, we have $\rho_{ls}=\rho_l$, $\rho_{gs}=\rho_g$. 

In addition to the aforementioned $e_{w1}$ formulation, there are other approaches utilized to constrain the dynamics of the contact line. One such method is the constant contact angle boundary condition, which enforces the dynamic contact angle to be equal to the equilibrium contact angle. The energy formulation for this condition is expressed as follows~\cite{ding2007wetting,chen2023diffuse}
\begin{equation}
    e_{w2}=\lambda\cot{\theta_{eq}}\int_{\rho_{gs}}^{\rho_{ls}}\frac{\partial \rho}{\partial x} \ d\rho +C.
\end{equation}
Another stress-balanced energy formulation \cite{qian2006variational}
\begin{equation}\label{w_en3}
    e_{w3}=-\frac{\sigma}{2}\cos{\theta_{eq}}\sin{\frac{\pi\phi}{2}},
\end{equation}
where $\phi=\frac{2\rho-\rho_l-\rho_g}{\rho_l-\rho_g}$ known as the order parameter changing from $\phi=[-1,1]$, and $\sigma$ is the surface tension between two fluids.
\begin{figure}
  \centering

  \includegraphics[width=\textwidth]{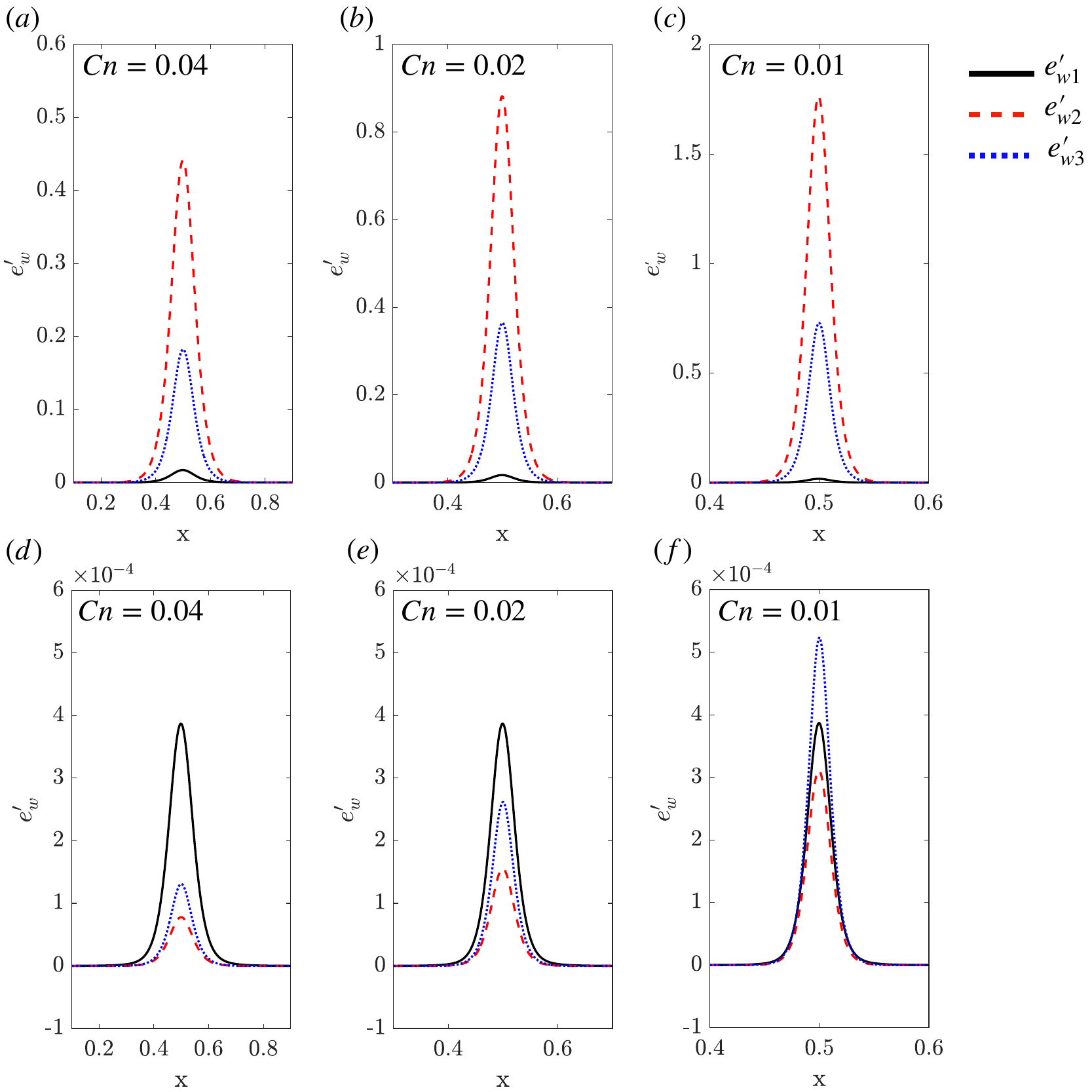}

\caption{Comparison of the wetting potential: $e_{w1}'$, $e_{w2}'$, and $e_{w3}'$ along the horizontal direction with initial $Cn=[0.01-0.04]$ for (a), (b), (c) with $\theta_{eq}=\frac{\pi}{12}$, and (d), (e), (f) with $\theta_{eq}=\frac{\pi}{4}$. \label{wallcp}}

\end{figure}
As well as a thermodynamic consistent formulation based on the pseudopotential lattice Boltzmann method can be shown as \cite{huang2022surface,huang2023equation}
\begin{equation}
    e_{w4}=-K_{EOS} K_{INT}\frac{\gamma(\rho_l-\rho_g)} {2\zeta} \tanh{\zeta\phi},
\end{equation}
where $K_{EOS}$, $K_{INT}$ are scaling factors to adjust the interface thickness of the phase field method, $\gamma$ and $\zeta$ are independent parameters that determine the contact angle.

Among the various wetting energy formulations, $e_{w2}$ maintains a constant contact angle throughout the evolution of the contact line. However, this approach may violate thermodynamic principles. Particularly, when the equilibrium contact angle $\theta_{eq}$ is close to 0 or $2\pi$, the simulation system becomes highly unstable. On the other hand, the formulation of $e_{w3}$ is derived by considering the stress balance and minimizing the free energy, thereby ensuring the preservation of correct thermodynamics \cite{qian2003molecular,qian2006variational}. In order to establish the exact relationship between $\sigma$ and $\lambda$, it is necessary to further determine the profile of the interface, as shown in the work by Chen et al.~\cite{chen2023diffuse}. This relationship plays a crucial role in ensuring the accuracy of the contact line dynamics. In Figure~\ref{wallcp}, we present a comparison of the wetting potential values along the interface for different Cahn numbers, denoted as $Cn=\delta/L$, where $\delta$ represents the initial interface thickness and $L$ is the length of the system. Specifically, in Figure~\ref{wallcp} (a), (b), and (c), we consider an equilibrium contact angle of $\theta_{eq}=\pi/12$ for varying values of $Cn$. It can be observed that, due to the large value of $\cot{\theta_{eq}}$ in the case of a small contact angle, the wetting potential of the energy $e_{w2}$ exhibits significantly higher values compared to the other two methods. Furthermore, when we increase the equilibrium contact angle to $\theta_{eq}=\pi/4$, as depicted in Figure~\ref{wallcp} (c), (d), and (e), the wetting energy formulation $e_{w3}$ exhibits varying peak values for different values of $Cn$. It is worth noting that the density profile is represented by a hyperbolic tangent function in each case. Consequently, the relationship between $\sigma$ and $\lambda$ is precisely determined as $\sigma\approx0.943\lambda/\delta$ when the value of $\delta$ is known.

The formulation of $e_{w4}$ is heavily influenced by the parameter selection and is more suitable for specific numerical methods. In recent studies, an implicit chemical potential boundary condition has been proposed to address the contact line problem \cite{wen2017chemical,yu2021alternative}. Due to the fully implicit nature of the method, it becomes challenging to accurately determine the contact angle precisely from the provided chemical potential value and temperature.

In our energy boundary condition, as described in Eq.(\ref{w_en1}), the computation of $e_{w1}$ through integration is required. However, in a realistic simulation, this value is not necessary. Therefore, this approach can be utilized as a general boundary condition that effectively preserves thermodynamic consistency. Additionally, the information regarding the interface thickness $\delta$ in $e_{w1}$ is implicitly incorporated into the bulk free energy, and all the essential parameters are computed locally. This approach successfully addresses the instability issues encountered in previous methods.

There are linear, quadratic, and cubic wetting energy formulations based on the C-H model. However, similar to the formulation $e_{w3}$, these formulations require prior relations to evaluate the interface thickness and determine the density profile on the boundary. Therefore, we have not considered these formulations in the current work. For a detailed analysis of these formulations, refer to \cite{briant2004lattice,liu2009wall}.

\section{Numerical Scheme}
\label{sec:numericalscheme}

To solve the governing equations presented in the previous section, we employ the two-step MacCormack methodology \cite{maccormack1982numerical,nadiga1995investigations}. To begin, we define a vector $\mathbf{f}$ consisting of the density $\rho$ and momentum $\rho \mathbf{u}$. Then, we proceed to reconstruct the governing equations using this vector Eqs.~(\ref{ns_c}), (\ref{ns_m}):
\begin{equation}
    \mathbf{f} = \begin{pmatrix} \rho \\ \rho\mathbf{u} \end{pmatrix}.
\end{equation}
Eqs.~(\ref{ns_c}), (\ref{ns_m}) can now be expressed as the functions of $\mathbf{f}$:
\begin{equation}\label{pre_cor}
    \partial_t \mathbf{f} +\nabla\cdot \mathbf{F} (\rho,\nabla \rho, \nabla^2\rho)=0,
\end{equation}
where $\mathbf{F}$ can be further expressed as:
\begin{equation}
    \mathbf{F} = \begin{pmatrix} \rho \mathbf{u} \\ \rho \mathbf{u}\otimes\mathbf{u}+p\mathbf{I}-\mathbf{\sigma}_{surf}-\mathbf{\sigma}_{vis}\end{pmatrix}.
\end{equation}

As shown in \cite{nadiga1995investigations}, Eq.~(\ref{pre_cor}) can be solved by a precondition and correction finite difference method. The time derivative is dealt with in a fully explicit manner: 
\begin{equation}
    \mathbf{f}^*=\mathbf{f}^n -\Delta t \nabla^{bck} \cdot \mathbf{F}^n,
\end{equation}
\begin{equation}
    \mathbf{f}^{n+1}=\frac{1}{2}\left(\mathbf{f}^n+\mathbf{f}^*\right) -
    \frac{\Delta t}{2} \nabla^{fwd} \cdot \mathbf{F}^*.
\end{equation}
Here, $\nabla^{fwd}$ stands for forward finite difference:
\begin{equation}\label{fwd}
    \nabla^{fwd}\phi(\mathbf{x})=\frac{\phi(\mathbf{x}+h)-\phi(\mathbf{x})}{h},
\end{equation}
$\nabla^{bck}$ is the backward finite difference:
\begin{equation}\label{bck}
    \nabla^{bck}\phi(\mathbf{x})=\frac{\phi(\mathbf{x})-\phi(\mathbf{x}-h)}{h},
\end{equation}
and $\nabla^{ctr}$ represents the central finite difference:
\begin{equation}\label{center}
   \nabla^{ctr}\phi(\mathbf{x})=\frac{\phi(\mathbf{x}+h)-\phi(\mathbf{x}-h)}{2h}.
\end{equation} 
In addition, the derivative computation appears in $\mathbf{F}$ can be computed by $\mathbf{F}^n(\rho^n,\nabla^{fwd}\rho^n,\nabla^2_{ctr}\rho^n)$, and $\mathbf{F}^*(\rho^*,\nabla^{bck}\rho^*,\nabla^2_{ctr}\rho^*)$ respectively.

Our simulation is implemented using the free code platform called Basilisk, which is a common tools language for the Octree structure utilizing adaptive mesh refinement methods \cite{popinet2003gerris, popinet2009accurate}. Given that our method relies on finite differences and is fully explicit, the strategy for adaptive mesh refinement is straightforward. The complete code is now accessible at the following link: http://basilisk.fr/sandbox/zchmacchiato/.
\section{Results}
\label{sec:results}
\begin{figure}
  \centering

  \includegraphics[width=0.8\textwidth]{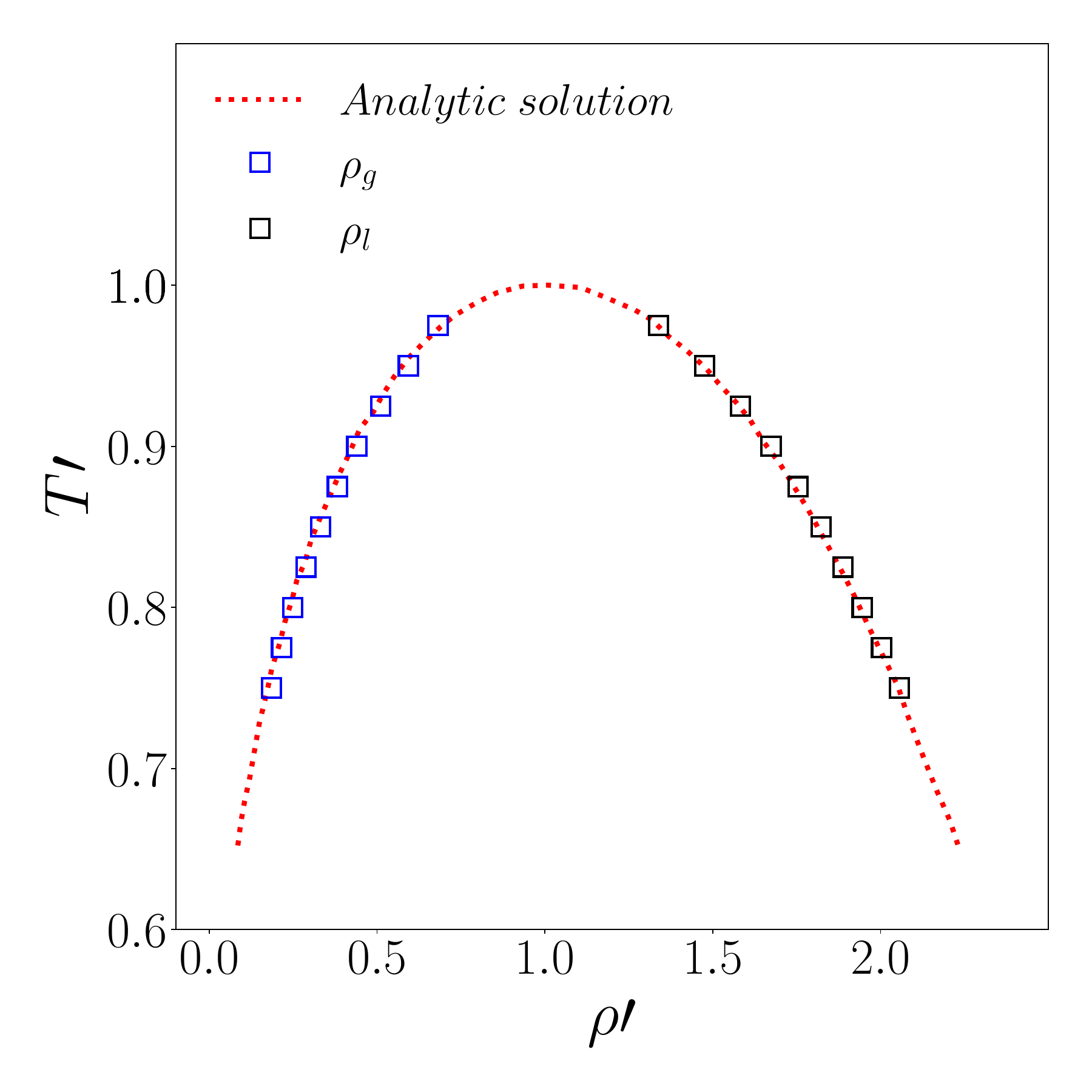}
    \caption{Equilibrium density values coexisting at different temperatures. The dashed line represents the analytic solutions derived from the Maxwell construction. The simulation results, indicated by square symbols, show the densities of the lighter gas ($\rho_g$) and the higher liquid ($\rho_l$) under various temperature conditions.\label{vdw_tem} }

\end{figure}

In this study, the VDW model is utilized to simulate phase transformations for an isothermal single species multi-phase system. The interface between the two phases undergoes changes from the initial shape to the equilibrium shape, resulting in fluid flow. Our simulations aim to assess the stability, energy oscillation, and morphology changes that occur during this phase transition process.

\subsection{single droplet simulation}
To validate the numerical method, we simulate the coexisting saturated density values at a fixed temperature. The simulation begins by initializing a single droplet with a radius of $r$ inside a gas tank, and it continues until the system reaches an equilibrium state. The initial density profile is represented by a hyperbolic tangent function:
\begin{equation}\label{single}
    \rho(\mathbf{x},0)=\frac{\rho_l+\rho_g}{2}-\frac{\rho_l-\rho_g}{2} \tanh{\frac{|\mathbf{x}-\mathbf{x_0}|-r}{\delta}}.
\end{equation}
Here, $|\mathbf{x}-\mathbf{x_0}|$ represents the distance between the local position and the droplet interface, and $\delta$ denotes the initial interface thickness, $r$ is the radius of the initial droplet. In Figure~\ref{vdw_tem}, we compare the simulation results of the density values with the analytic solutions obtained from the Maxwell construction. Our numerical scheme accurately captures the results, which are in good agreement with the theoretical solutions.
\begin{figure}
  \centering

  \includegraphics[width=\textwidth]{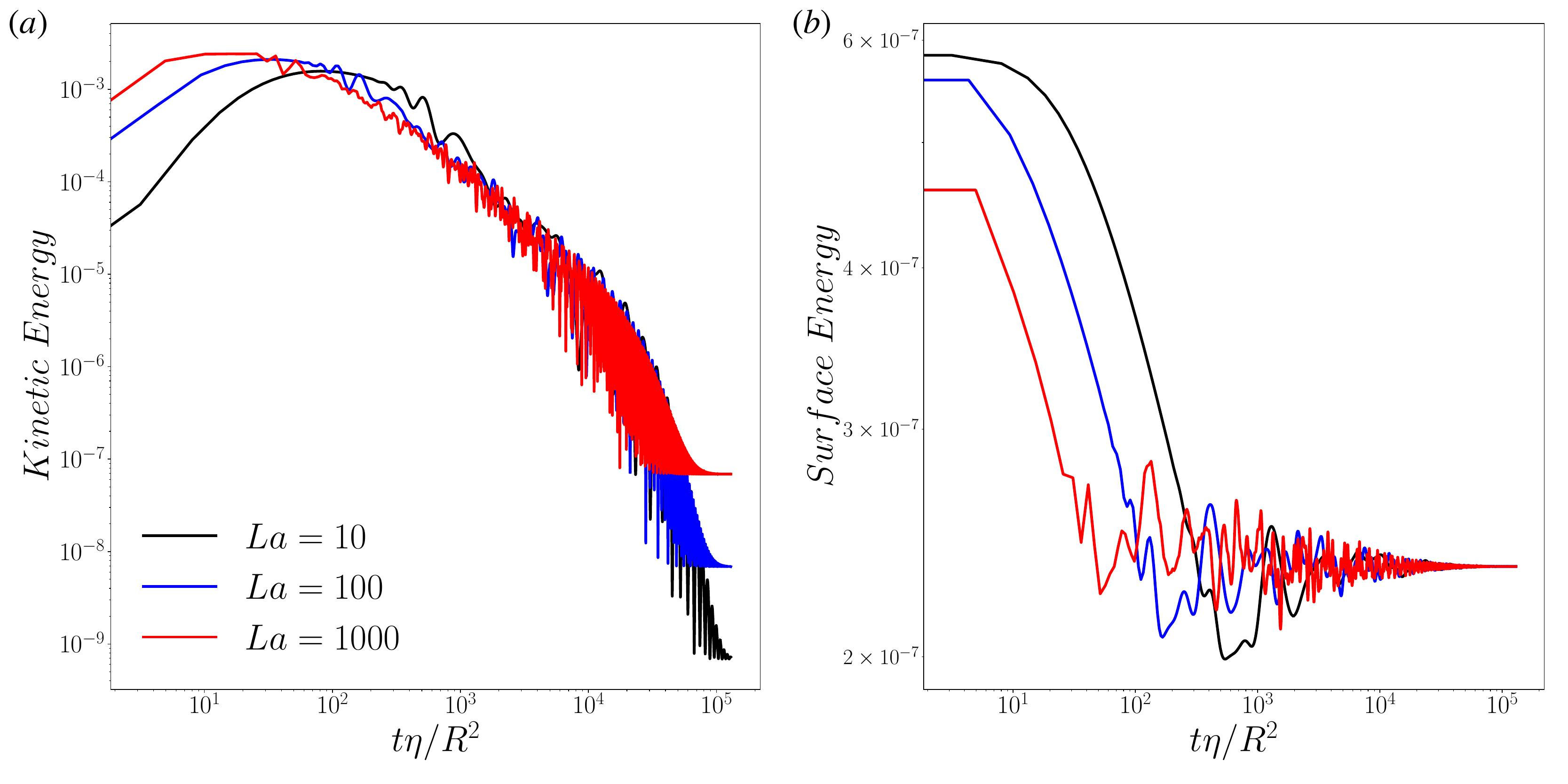}
   \caption{Logarithmic evolution of (a) kinetic energy $E_k$ and (b) surface energy $E_s$ over the domain for the single static droplet simulation. \label{energy}}

\end{figure}

We proceed with the simulation of a single droplet in a square domain under a constant temperature $T'=0.95$ with periodic boundaries. Based on the results shown in Figure~\ref{vdw_tem}, the approximate saturated density of the liquid phase is $\rho_l\approx1.46$, and that of the gas phase is $\rho_g\approx0.58$. In this test, we do not consider the viscosity ratio. It is important to note that the exact density profile in a static solution is more complex, but it can be qualitatively represented by a hyperbolic tangent function as given in Eq.(\ref{single}). Consequently, the presence of different initial density values compared to the saturated density values introduces oscillations in the simulation. The initial density distribution, along with the surface tension stress, drives the droplet towards its equilibrium shape, while pressure helps separate the saturated density profile simultaneously. In an ideal scenario, with sufficient evolution time, we would expect a constant surface energy $E_s=\int_\Omega e_s$ and zero kinetic energy $E_k=0$. However, due to unbalanced numerical schemes and the choice of the surface force formulation\cite{lee2006eliminating}, spurious currents can occur.

In this test, we characterize the system using the Laplace number, $La=\lambda \rho_c R/\eta^2$, where $R$ is the initial radius of the droplet. With $La\gg1$, we expect a significant surface effect that induces pronounced spurious currents when the system reaches equilibrium~\cite{popinet1999front}. The logarithmic evolution of kinetic energy and surface energy is presented in Figure~\ref{energy}. We vary $La$ from $10$ to $1000$. As the viscous force dissipates the system's energy and balances the oscillations caused by capillary waves, reducing $La$ leads to a rapid decrease in kinetic energy. The viscous dissipation gradually consumes the energy associated with the droplet shape, causing the kinetic energy to converge to a small, constant value. In our simulations, the final kinetic energy, attributed to spurious currents, does not reach zero. However, the surface energy converges to the same value for different $La$ values, indicating that the surface effect accelerates the system's attainment of the equilibrium profile. When $La\geq1000$, oscillations in the energies are observed. In such high-temperature systems, the significant surface effect induces capillary waves around the phase interface. The imbalance between surface tension and thermodynamic pressure, combined with the explicit numerical scheme, leads to the generation of spurious currents, preventing the system from reaching zero kinetic energy.
\subsection{Dynamics of liquid-vapor separation}

\begin{figure}
  \centering

  \includegraphics[width=\textwidth]{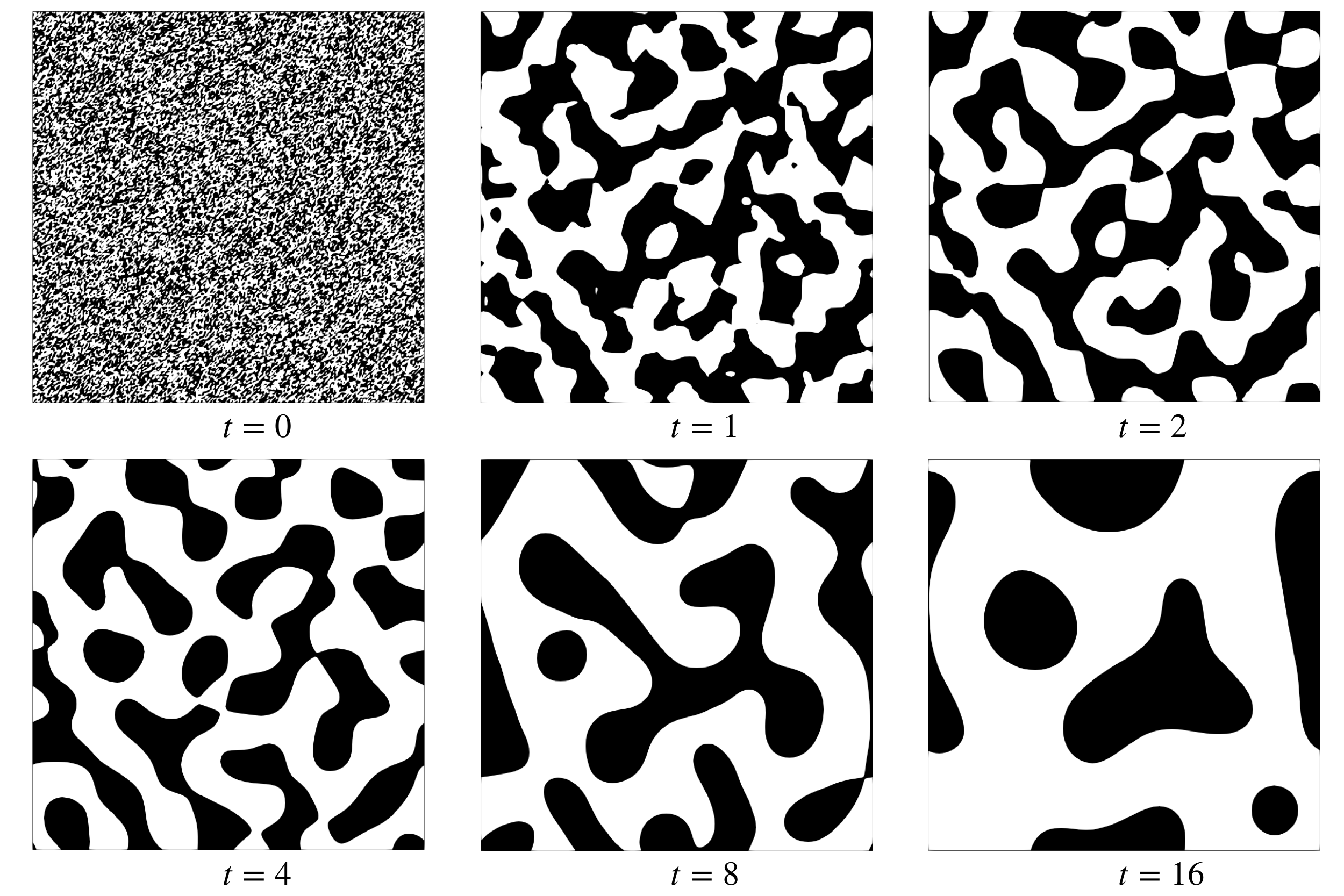}
\caption{Temporal evolution of the dynamics of liquid-vapor separation for $t = [0, 16]$. \label{spin}}

\end{figure}

\begin{figure}
  \centering

  \includegraphics[width=0.6\textwidth]{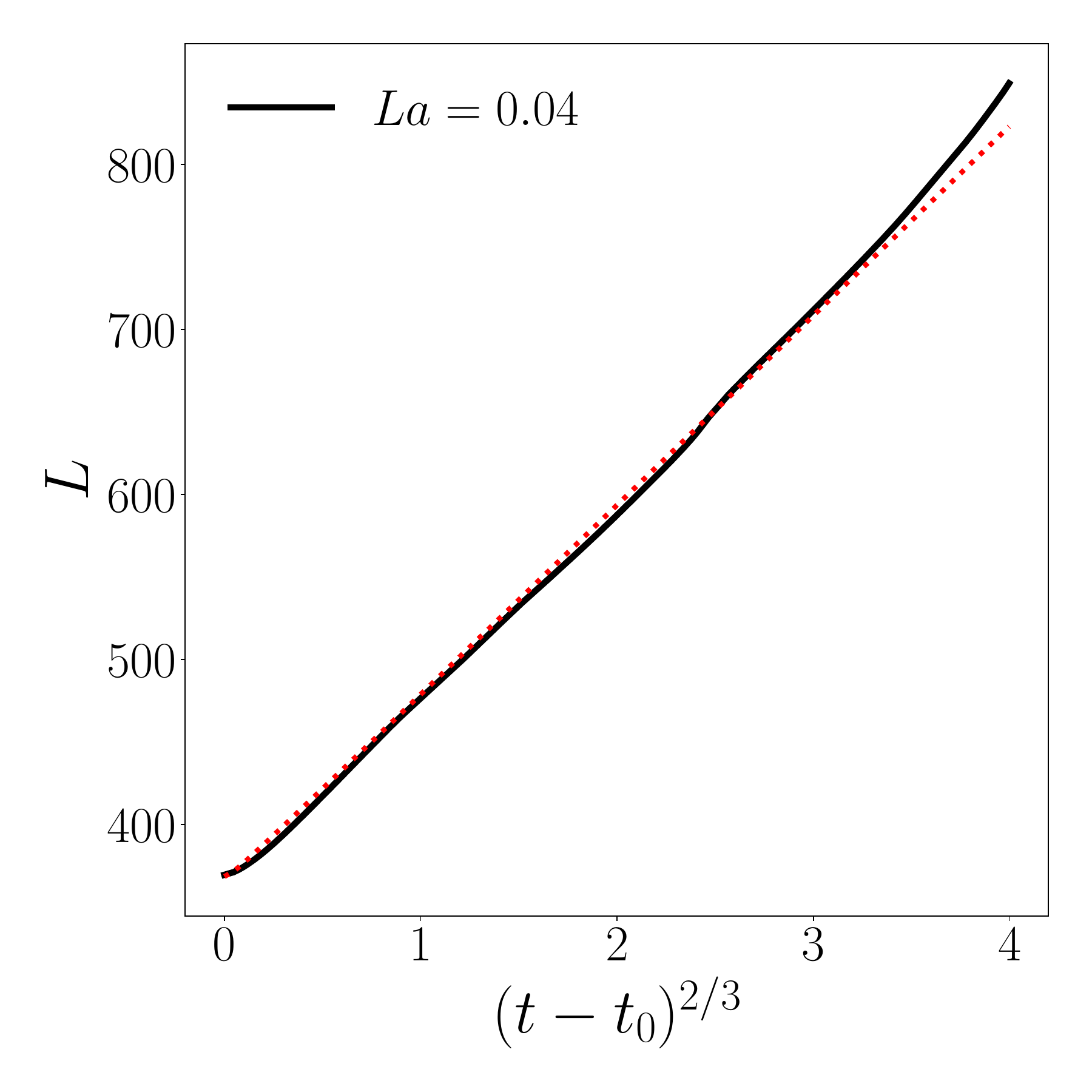}
\caption{Evolution of domain length $L$ versus time $(t-t_0)^{2/3}$. The red dashed line represents the reference curve for theoretical solutions. \label{sc}}

\end{figure}

In this section, we explore the applicability of the VDW model to the dynamics of liquid-vapor separation, aiming to assess its performance in a complex morphology-changing problem. Additionally, we have incorporated adaptive mesh refinement into the simulation to enhance efficiency. When a single species is subjected to a temperature close to the critical temperature and possesses a random density profile, the pressure and surface stress act as driving forces, leading the mixture to undergo phase separation. This results in coarsening dynamics and the formation of two distinct phases: one with a higher density and the other with a lower density. In the VDW model, the phase separation solely relies on the equilibrium density corresponding to specific temperatures. This enables the system to minimize the free energy during its evolution. 

As depicted in Figure~\ref{spin} (a), we initiate the simulation by introducing a single species with a random density fluctuation within a 2D square domain. The boundaries of the domain are set as periodic conditions to ensure continuity. The initial density profile is defined as follows:
\begin{equation}
        \rho(\mathbf{x},0)=\rho_c+0.2\rho_c (rand),
\end{equation}
where the amplitude of random fluctuation is set to $0.2 \rho_c$. The random number for generating the fluctuations is obtained from the random seed $rand=[-1,1]$. The phase separation is characterized by the growth of the domain length scale, defined as $L=L_0^2/\chi_m$, where $L_0^2$ represents the area of the square domain, and $\chi_m=\langle C^2(1-C)^2\rangle$ is the space average quantity parameter associated with the concentration of the gas phase, denoted as $C=(\rho-\rho_g)/(\rho_l-\rho_g)$ \cite{semprebon2016ternary}. In our previous work, we utilized the explicit method to investigate the dynamics of liquid-vapor separation under constant temperature conditions \cite{nadiga1995investigations}. When the system temperature was set to $T'=0.85$, simulation results exhibited a growth rate characterized by $L\sim(t-t_0)^{0.7}$, which was close to but slightly higher than the $(t-t_0)^{2/3}$ growth rate reported by Miranda et al.~\cite{miranda2006spinodal}. In the present study, we simulate the dynamics of liquid-vapor separation under $T'=0.95$ with a Laplace number of $La=0.04$. In this example, the simulations are performed with adaptive meshes using the feature of Basilisk~\cite{popinet2003gerris}. The smallest (dimensionless) cell size, $\Delta$, used is 0.0039 in order to fully resolve the liquid-vapor interface.   The results presented here are obtained by averaging over 5 runs with different random initial density configurations.

The evolution of the mixture's morphology at different time steps is shown in Figure~\ref{spin}. Over time, the complexity of the mixture gradually diminishes, and the influence of surface tension becomes prominent, resulting in the formation of circular liquid droplets in the later stages. In Figure~\ref{sc}, we compare the simulation results of the domain length scale $L$ evolution when $La=0.04$ with the corresponding theoretical solutions depicted by the red dashed curve. It can be observed that our simulation results exhibit a close agreement with the theoretical prediction $L\sim (t-t_0)^{2/3}$.

\subsection{Equilibrium contact angle and energy evolution}
\begin{figure}
  \centering

  \includegraphics[width=1.1\textwidth]{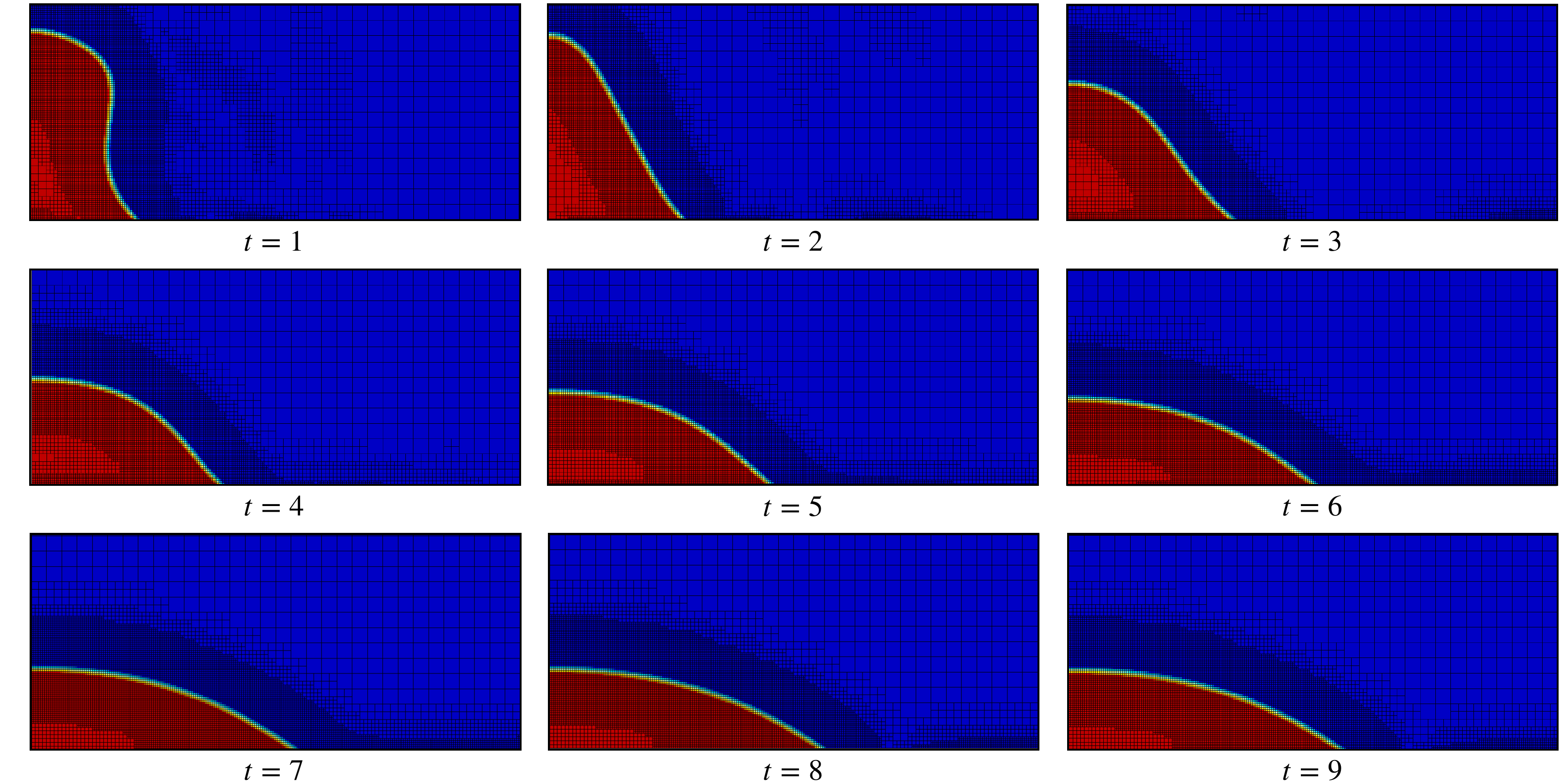}

\caption{Temporal evolution of the contact line with equilibrium contact line $\theta_{eq}=\frac{\pi}{6}$ over time range $t=[1,9]$. The black lattices indicate the grids. The resolution of the simulation is dynamically tuned between fine and coarse levels based on the density profile.\label{CA30}}

\end{figure}
\begin{figure}
  \centering

  \includegraphics[width=\textwidth]{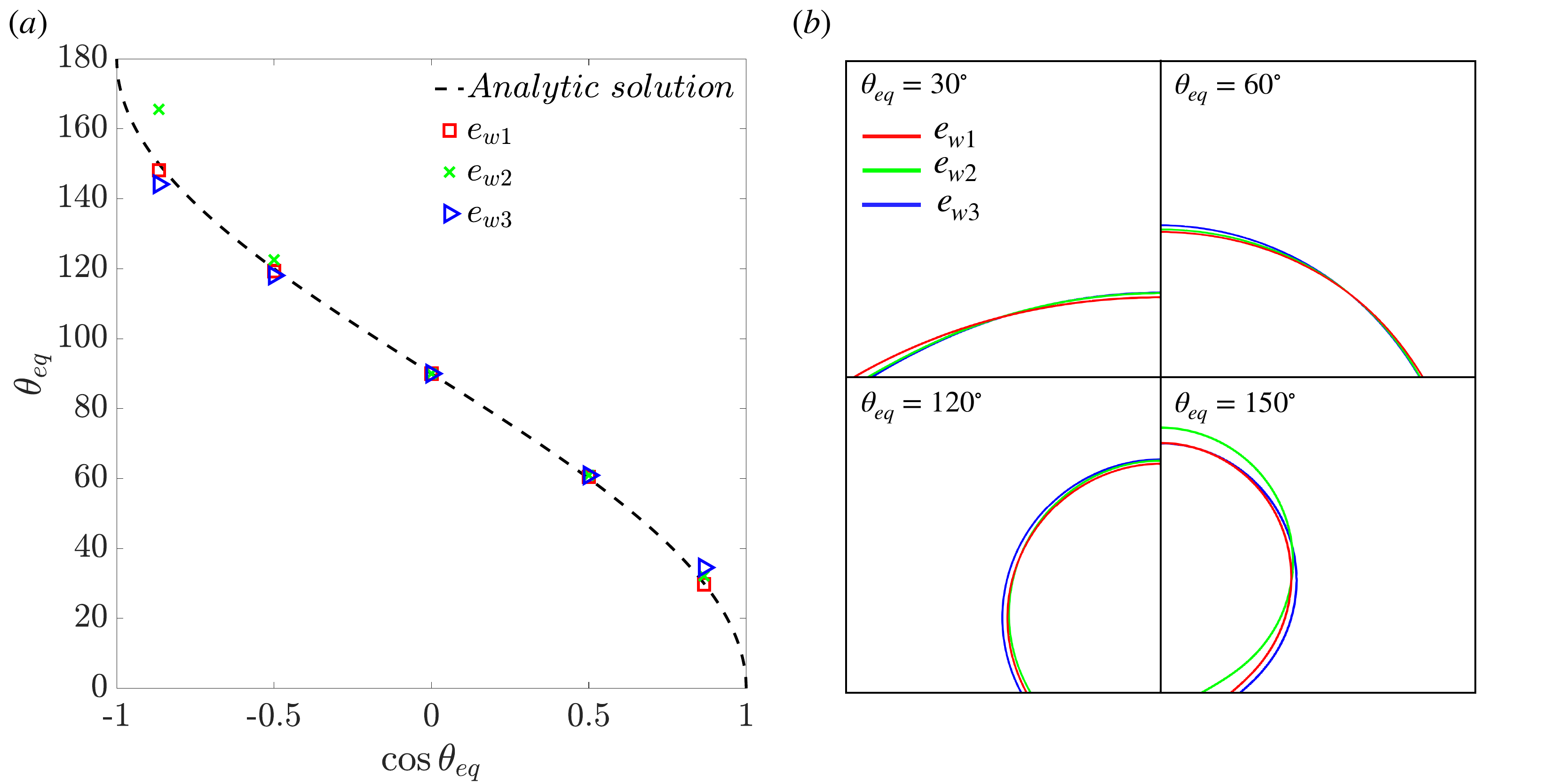}

\caption{(a) Comparison between contact angle simulation and the analytic solution $\theta_{eq}$ for $\theta_{eq}=[\frac{\pi}{3}-\frac{5\pi}{6}]$ with different wetting potentials. (b) Simulation comparison between different wetting potential forms. \label{CA}}

\end{figure}

In the previous section, we compared various boundary condition methods based on their profiles along the interface for the 1D planar case. Now, we employ different boundary conditions to model the equilibrium contact angle and assess the energy evolution of the contact angle simulation.
\begin{figure}
  \centering

  \includegraphics[width=\textwidth]{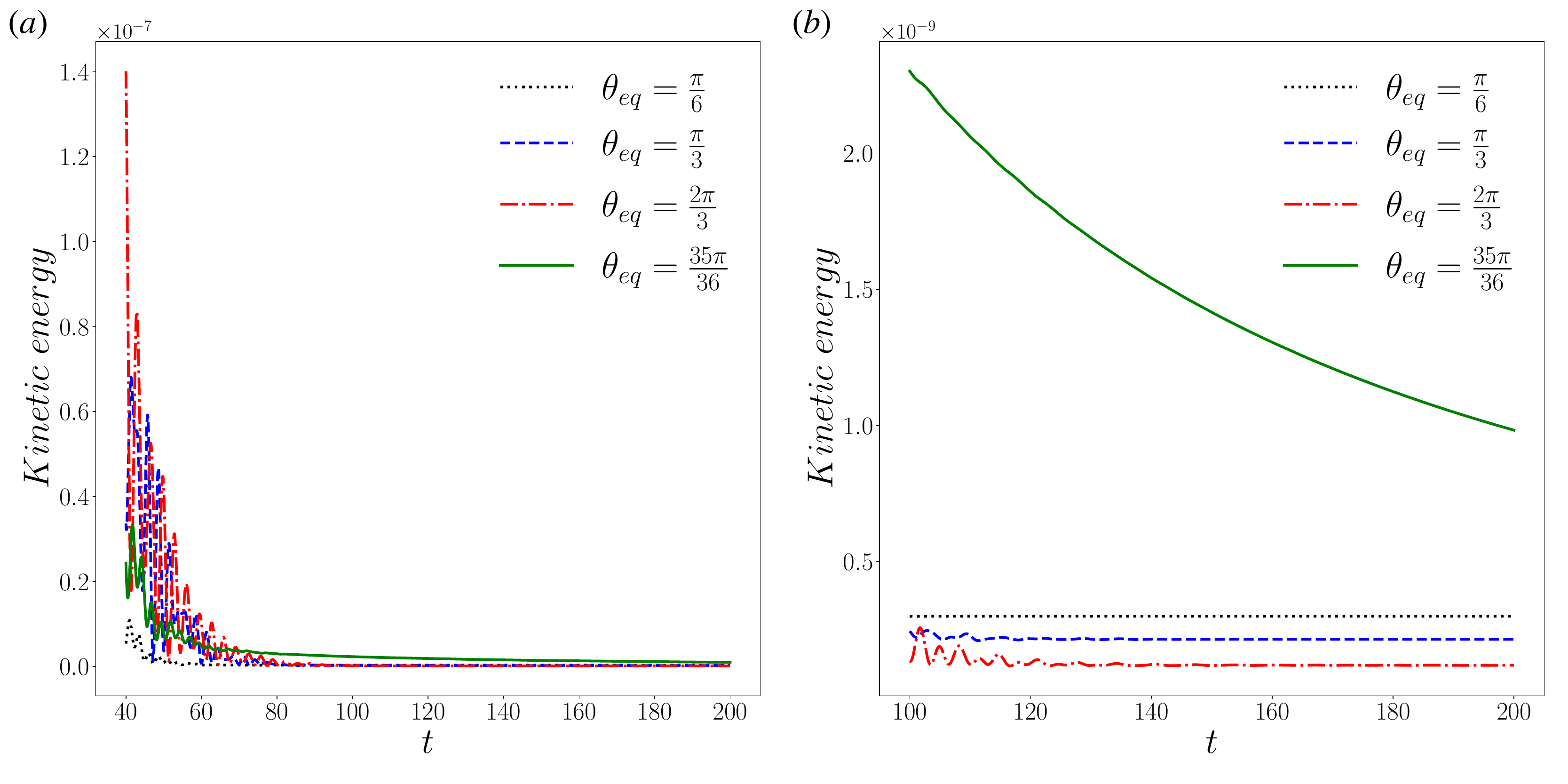}

\caption{Kinetic energy evolution of $\theta_{eq}=[\pi/6-35\pi/36]$ for $e_{w1}$ during (a) $t=[40-200]$ and (b) $t=[100-200]$. \label{caenergy}}
\end{figure}

For this test, we start with a liquid droplet of radius $r=0.2L$, where the density is set to $\rho=\rho_l$, located on a solid boundary. The region to the left of the droplet is filled with gas, with a density of $\rho=\rho_g$. The density profile function and the velocity can be defined as follows:
\begin{equation}\label{droplet}
    \rho(\mathbf{x},0)=\frac{\rho_l+\rho_g}{2}-\frac{\rho_l-\rho_g}{2}\tanh{\frac{|\mathbf{x}-\mathbf{x_0}|-r}{\delta}},
\end{equation}

\begin{equation}\label{velo}
    \mathbf{u}(\mathbf{x},0)=\mathbf{0}.
\end{equation}
Here, $|\mathbf{x}-\mathbf{x_0}|$ represents the distance to the interface of the liquid droplet, and $\mathbf{x_0}$ denotes the center position of the droplet. The temperature for this simulation is fixed at $T'=0.95$. The viscosity $\eta$ and surface energy coefficient $\lambda$ are chosen to satisfy $La=4$ for all simulations. To ensure stable simulations, the time interval $\Delta t$ needs to satisfy $\frac{\eta\Delta t}{\Delta x^2}\leq0.1$. Additionally, the Courant-Friedrichs-Lewy (CFL) condition is imposed with $CFL=\frac{|\mathbf{u}{\text{max}}|\Delta t}{\Delta x}\leq0.1$. Since the equilibrium interface thickness, $\delta_{eq}$, is unknown in the $e_{w3}$ formulation, we set $\delta_{eq}=0.3$ and approximate $\sigma$ as $\approx3.143\lambda$.
\begin{figure}
  \centering

  \includegraphics[width=\textwidth]{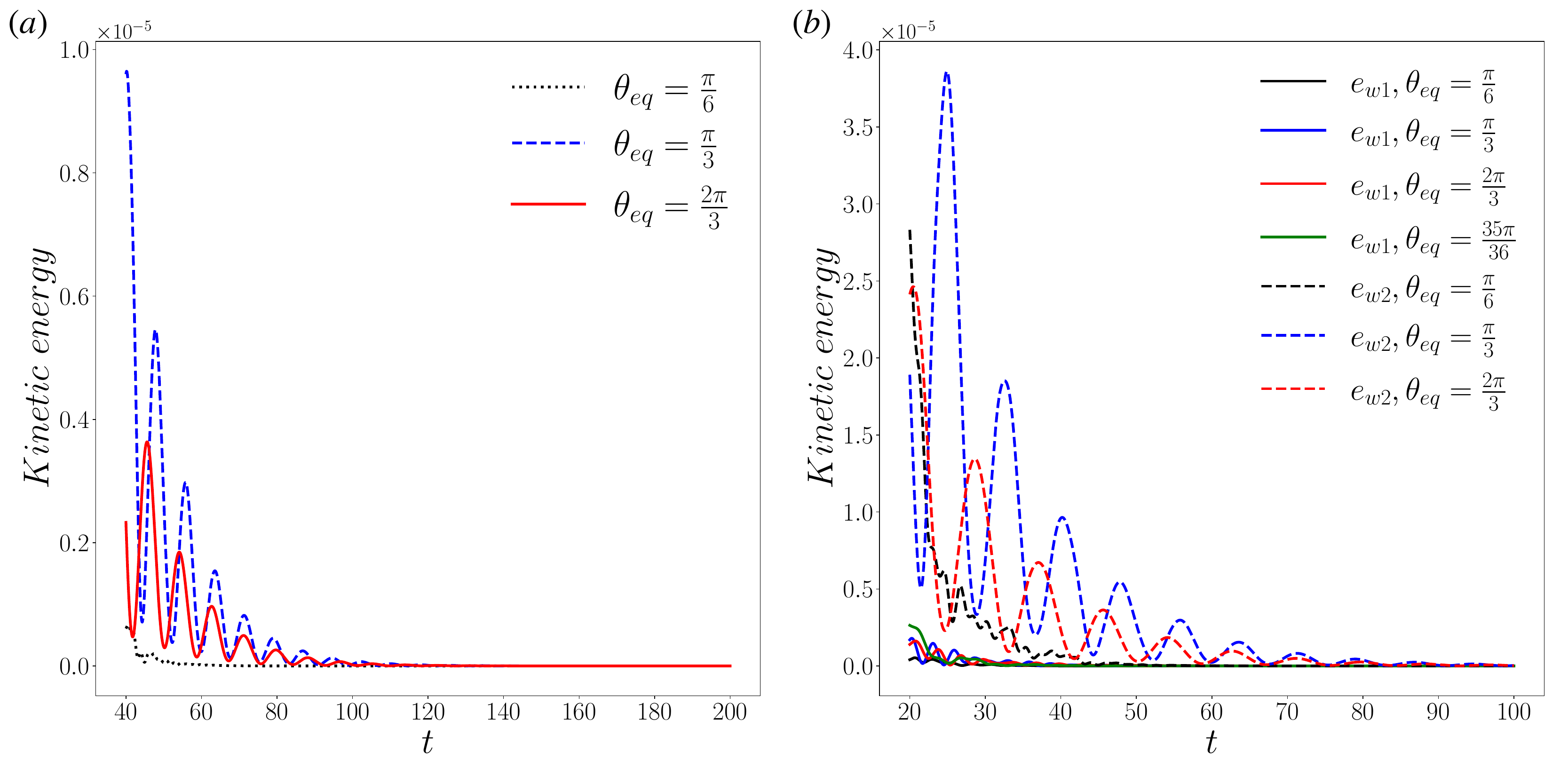}
\caption{(a) Kinetic energy evolution of $\theta_{eq}=[\pi/6-2\pi/3]$ for $e_{w2}$. (b) Comparison of kinetic energy evolution between $e_{w1}$ and $e_{w2}$ for $\theta_{eq}=[\pi/6-2\pi/3]$ during $t=[20-100]$. \label{cacomp}}

\end{figure}
\begin{figure}
  \centering

  \includegraphics[width=\textwidth]{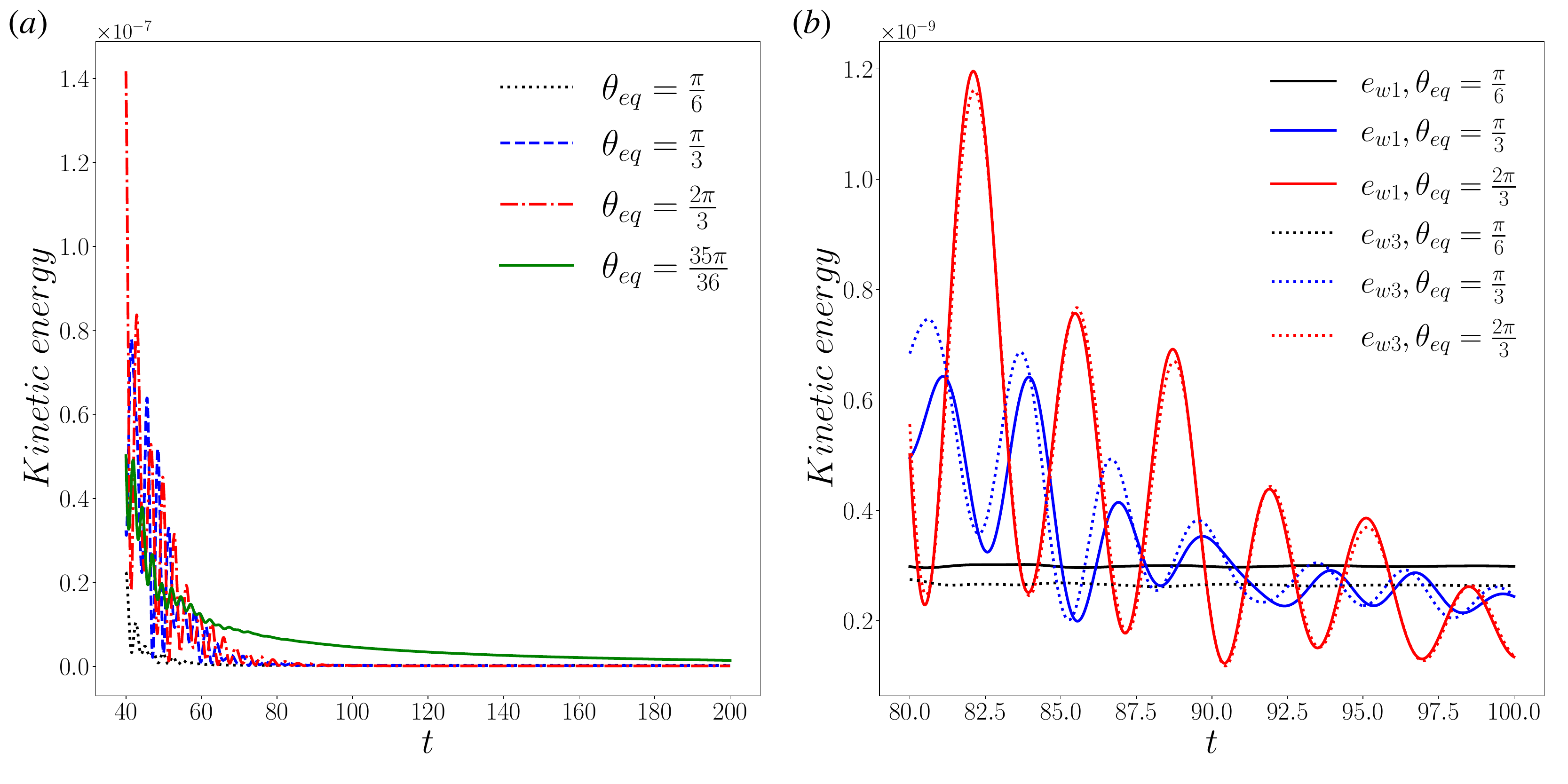}

\caption{(a) Kinetic energy evolution of $\theta_{eq}=[\pi/6-35\pi/36]$ for $e_{w3}$. (b) Comparison of kinetic energy evolution between $e_{w1}$ and $e_{w3}$ for $\theta_{eq}=[\pi/6-2\pi/3]$ during $t=[80-100]$. \label{stresscom}}

\end{figure}

The equilibrium contact angle for each simulation is computed using the method described in \cite{zhao2023interaction}. The simulation results are evaluated when the kinetic energy reaches a constant value, typically when $t\gg1$. In the provided figures, the interface position is defined as $\rho=\rho_c$.

Figure~\ref{CA30} illustrates the evolution of density profiles for an equilibrium contact angle of $\theta_{eq}=\pi/6$ at various time points. In this study, we employ adaptive mesh refinement techniques, where the resolution of the mesh is determined by the density distribution. Notably, the grids demonstrate a clear refinement in the contact line region as the droplet spreads over the solid surface. Figure~\ref{CA}(a) compares three different boundary conditions with the corresponding analytic solution, indicated by the black dashed line. The equilibrium shapes of the droplets residing on solids with different contact angles are shown in Figure~\ref{CA}(b). Upon comparison, it can be observed that when the equilibrium contact angle $\theta_{eq}$ is set to $\pi/2$, the results from the three methods align well with each other. Similarly, when the equilibrium contact angle approaches $\pi/2$, the simulation results of the form $e_{w1}$ exhibit good agreement with the reference curve. However, when $\theta_{eq}$ is set to $2\pi/3$ and $\pi/6$, the results from form $e_{w2}$ deviate from the analytic solutions. Additionally, the accuracy of the results from form $e_{w3}$ is lower compared to those from $e_{w1}$ or the analytic solutions. By considering these comparisons, we can conclude that the form $e_{w1}$ consistently provides accurate results across a wider range of contact angles compared to the other two formulations.

It is important to evaluate the energy evolution of the contact line moving until the system reaches the equilibrium state, as it provides insights into the contact line dynamics of the system \cite{huang2022surface}. Figure~\ref{caenergy} illustrates the evolution of kinetic energy during the simulation. Once the droplet achieves its equilibrium shape, it stops evolving, and the kinetic energy initially increases and then gradually decreases toward zero. For the energy form $e_{w1}$, in the late stages of the simulation, the kinetic energy of each case converges to a very small value, approximately $E_{k}\sim10^{-10}$. To further analyze the differences in kinetic energy evolution, Figure~\ref{cacomp} compares the kinetic energy profiles for the energy forms $e_{w1}$ and $e_{w2}$. It is worth noting that for contact angles $\theta_{eq}>5\pi/6$, the simulation process becomes highly unstable when using form $e_{w2}$. Hence, the results for an even larger contact angle, $\theta_{eq}=35\pi/36$, are not compared. The comparison in Figure~\ref{cacomp}(b) clearly demonstrates that the kinetic energy evolution differs significantly between the two boundary conditions. This indicates that the contact line dynamics associated with these two methods during the simulation are also distinct. Moreover, when using the $e_{w1}$ formulation for the boundary, the system reaches the equilibrium state more rapidly. In summary, the analysis of the kinetic energy evolution supports the superiority of the $e_{w1}$ formulation, as it leads to faster convergence to the equilibrium state and provides more stable contact line dynamics compared to the $e_{w2}$ formulation.

We proceed to compare the energy forms $e_{w1}$ and $e_{w3}$ in Figure~\ref{stresscom}. In order to maintain consistency between the two boundary conditions, the interface thickness $\delta_{eq}=0.164$ is obtained from the equilibrium state of the simulation using $e_{w1}$ as the wetting energy. The comparison in Figure~\ref{stresscom}(b) reveals that the kinetic energy evolution of both cases is qualitatively consistent, and even the capillary-induced oscillation exhibits similar characteristics. It is worth noting that the idea behind form $e_{w3}$ stems from the concept of stress balance between the gas phase and liquid phase at the interface region \cite{qian2003molecular,qian2006variational,carlson2009modeling,carlson2012universality,carlson2011dissipation}. The surface tension difference $\Delta \sigma$ between the solid-gas and solid-liquid interfaces remains continuous along the boundary, and the dynamics of the moving contact line simulated using this method have been widely employed and compared with molecular dynamics and experimental studies \cite{qian2006variational,carlson2009modeling,carlson2012universality}. Consequently, from a kinetic energy and dynamics standpoint, similar outcomes can be achieved whether we utilize $e_{w1}$ or $e_{w3}$ as the wetting energy. This suggests that both formulations can capture the essential features of the contact line dynamics and yield comparable results in kinetic energy and capillary-driven oscillations.

\section{Concluding Remarks}

In this study, we investigated an explicit finite difference method for solving the Van der Waals (VDW) multi-phase flow. Based on the MacCormack methodology, the numerical scheme provided qualitative simulation results for single static droplets and the dynamics of liquid-vapor separation. We proposed a general energy-based approach to address the contact line problem by relating the wetting energy to bulk free energy and surface energy, and compared them with existing boundary condition methods.

In the simulation tests, we evaluated the energy evolution and spurious currents of single static droplets under different Laplace numbers ($La$). As $La$ was decreased to approximately 1, we observed a more stable equilibrium system with reduced intensity of spurious currents. We validated our method by analyzing the growth of domain length during the liquid-vapor separation process, and our results were in good agreement with the predicted solution $L=(t-t_0)^{2/3}$. Using the general energy-based boundary condition, we achieved highly consistent equilibrium contact angles with the predicted analytic solution. However, the other two existing methods failed to provide qualitative results due to large wetting potential and uncertain interface thickness.

Furthermore, the kinetic energy of the simulation for the equilibrium shape of the sessile droplet converged to $E_k\sim10^{-10}$, which is at a similar level as the simulation of spurious currents in the single static droplet. Additionally, we observed consistent dynamics between the energy-consistent boundary condition and the stress balance boundary condition when the same interface thickness was employed in both approaches.

Overall, our study demonstrated the effectiveness of the explicit finite difference method for VDW multi-phase flow and provided valuable insights into the energy-based approach for modeling contact line phenomena.


\section*{Appendix}\label{ap}
We here establish the distinction between the Korteweg stress-based surface force and the potential-based surface force. We begin by examining the force terms of the momentum equation, Eq.~(\ref{ns_m}), excluding the contribution from viscous dissipation
\begin{equation}\label{flux_stress1}
  \nabla\cdot ( \sigma_s - p\mathbf{I} ) =    \nabla\cdot\left(\lambda\left[\left( \frac{1}{2}|\nabla\rho|^2+\rho\nabla^2\rho\right)\mathbf{I}-\nabla\rho\otimes\nabla\rho\right]-p\mathbf{I}\right).
\end{equation}
In this context, the pressure is determined by the equation of state, which can be obtained from the thermodynamic energy as $p=\rho^2\partial f_0/\partial \rho$, as explained in the main text. By substituting this expression into Eq.~(\ref{flux_stress1}), we obtain the following result:
\begin{equation}\label{flux_stress2}
    \nabla\cdot ( \sigma_s - p\mathbf{I} ) =  \nabla \left(\frac{\lambda}{2}|\nabla\rho|^2+\lambda\rho\nabla^2\rho-\rho^2\frac{\partial f_0}{\partial \rho}\right) -\nabla\cdot\left(\lambda\nabla\rho\otimes\nabla\rho\right).
\end{equation}
In addition, the flux term of the momentum equations can be also represented by a potential form surface force~\cite{jacqmin1996energy}
\begin{equation}\label{flux_potential1}
    -\rho\nabla\mu_{mix}=-\nabla\rho\mu_{mix}+\mu_{mix}\nabla\rho,
\end{equation}
with mixed chemical potential $\mu_{mix}=\partial (\rho f_0)/\partial\rho-\lambda\nabla^2\rho$. Eq.~(\ref{flux_potential1}) can be further simplified to
\begin{equation}\label{flux_potential2}
    -\rho\nabla\mu_{mix}=\nabla \left(\lambda\rho\nabla^2\rho-\rho^2\frac{\partial f_0}{\partial \rho}\right)-\lambda\nabla^2\rho\nabla\rho.
\end{equation}
Finally, the additional stress term $\sigma_\rho$ can be evaluated by the difference between Eq.~(\ref{flux_stress2}) and Eq.~(\ref{flux_potential2})
\begin{equation}\label{flux_difference}
    \nabla\cdot \sigma_\rho=\nabla\cdot(\sigma_s-p\mathbf{I})+\rho\nabla\mu_{mix}
    =\nabla\cdot\left(\frac{\lambda}{2}|\nabla\rho|^2\mathbf{I}-\lambda\nabla\rho\otimes\nabla\rho\right)+\lambda\nabla^2\rho\nabla\rho,
\end{equation}
which can be simplified as
\begin{equation}\label{flux_difference2}
    \sigma_\rho=\lambda\left(|\nabla\rho|^2\mathbf{I}-\nabla\rho\otimes\nabla\rho\right)+C,
\end{equation}
The constant $C$ is typically set to zero in practice, as only the divergence of the stress term appears in the momentum equation. The additional stress term is mostly implicitly incorporated into the pressure term. In the case of a 1-D simulation, this term simplifies to zero.


 \bibliographystyle{elsarticle-num} 
 \bibliography{cas-refs}





\end{document}